\documentclass{article}

\usepackage{arxiv}

\usepackage[utf8]{inputenc} % allow utf-8 input
\usepackage[T1]{fontenc}    % use 8-bit T1 fonts
\usepackage{hyperref}       % hyperlinks
\usepackage{url}            % simple URL typesetting
\usepackage{booktabs}       % professional-quality tables
\usepackage{amsfonts}       % blackboard math symbols
\usepackage{nicefrac}       % compact symbols for 1/2, etc.
\usepackage{microtype}      % microtypography
\usepackage{lipsum}		% Can be removed after putting your text content
\usepackage{graphicx}
\usepackage{natbib}
\usepackage{doi}

\usepackage{blindtext}
\usepackage{color}
\usepackage{graphicx}
\usepackage{amsthm}
\usepackage{mathtools}
\usepackage{amsfonts}
\usepackage{amssymb,amsmath}
\usepackage{mathtools}
\usepackage{subcaption}
\usepackage{caption}
\usepackage{amsmath}
\usepackage{tcolorbox}
\usepackage{titlesec}
\usepackage{stackengine}
\usepackage{caption}
\usepackage{subcaption}

\usepackage{tikz}
\usepackage{calrsfs}
\DeclareMathAlphabet{\pazocal}{OMS}{zplm}{m}{n}

\usepackage{thmtools, thm-restate}

\title{Chore Cutting: Envy and Truth}

\author{ \href{https://orcid.org/0000-0001-5030-9645}{\includegraphics[scale=0.06]{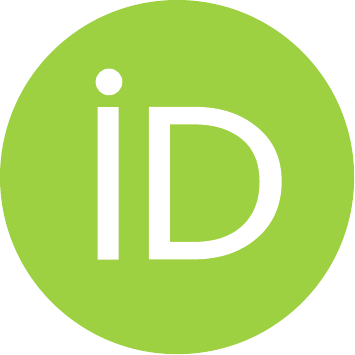}\hspace{1mm}Mohammad Azharuddin Sanpui} \\
	Department of Mathematics\\
	Indian Institute of Technology Kharagpur\\
	 India\\
	\texttt{azharuddinsanpui123@gmail.com} \\
}

\date{}

\hypersetup{
pdftitle={A template for the arxiv style},
pdfsubject={q-bio.NC, q-bio.QM},
pdfauthor={David S.~Hippocampus, Elias D.~Striatum},
pdfkeywords={First keyword, Second keyword, More},
}

\begin{document}
\maketitle

\begin{abstract}
We study the fair division of divisible bad resources with strategic agents who can manipulate their private information to get a better allocation. Within certain constraints, we are particularly interested in whether truthful envy-free mechanisms exist over piecewise constant valuations. We demonstrate that no deterministic  truthful  envy-free mechanism can exist in the connected-piece scenario, and the same impossibility result occurs for hungry agents. We also show that no deterministic, truthful dictatorship mechanism can satisfy the envy-free criterion, and the same result remains true for non-wasteful constraints rather than dictatorship. We further address several related problems and directions.
\end{abstract}

% keywords can be removed
\section{Introduction}
The problem of allocating a heterogeneous, divisible resource among a set of $n$ agents with varying preferences is essentially described as the problem of cutting a cake in intuitive concepts. The cake cutting problem is a fundamental topic in the theory of fair division  \cite{cakecuttingdispute,Cake,Fairdivisionandcollectivewelfare,HandbookCSC} and it has received a significant amount of attention in the domains of mathematics, economics, political science, and computer science \cite{TheEfficiencyoffairdivision,cakecuttingreallyisnotapieceofcake,feeiciencyoffairdivisionwithconnectedpiece,ChildrenCryingatbirthdayparties,cakecuttingnotjustchildplay,TheQuerycomplexityofcakecutting}. Dividing a cake fairly among agents is a challenging task.
\textit{Envy-freeness} and \textit{proportionality} are the most `important criteria of a fair allocation in the cake-cutting literature. In an envy-free allocation, every agent is pleased with the pieces they are allocated as opposed to any other agent's allocation. In a proportional allocation, each agent receives at least $\frac{1}{n}$ of the value he estimates to the cake. When all of the cake has been divided, envy-freeness entails proportionality. It is generally known that envy-free allocations always exists \cite{ANENVY-FREECAKEDIVISIONPROTOCOL} and even if we specify that each agent must receive a connected piece \cite{rental,Stromquist1980HowTC}. In addition to the existence, the algorithmic design aspect of the process has also been thought about for a long time \cite{ADiscreteandBoundedEnvy-FreeCakeCuttingProtocolforAnyNumberofAgents,Aboundedandenvy-freecakecuttingalgorithm,Adiscreteandboundedenvyfreecakecuttingprotocolforfouragents,Howtocutacakefairly,ANoteOnCakeCutting,Stromquist2008EnvyFreeCD}. For any number of agents, we are able to calculate a proportional allocation \cite{Howtocutacakefairly,ANoteOnCakeCutting} as well as an envy-free allocation \cite{ADiscreteandBoundedEnvy-FreeCakeCuttingProtocolforAnyNumberofAgents}.\newline
\\
In contrast, the dual problem of cake-cutting, also known as chore-cutting, seeks to allocate an undesirable resource to a set of $n$ agents, with each agent wishing to receive as little of the resource as possible. Chore division might model the allocation of chores within a household, liabilities in a bankrupt company, etc. Similar to the cake-cutting problem, dividing a chore fairly is also a challenging problem. The most important criteria of a fair allocation, similar to cake cutting, are \textit{envy-freeness} and \textit{proportionality}. For most questions in cake cutting there exist parallel questions in chore division; for instance, an $n$-person envy-free cake cutting algorithm, in which each agent is satisfied that no other agent has received a bigger piece in the their estimation, was found by Brams and Taylor \cite{ANENVY-FREECAKEDIVISIONPROTOCOL}, while the equivalent chore-cutting result found by Peterson and Su \cite{Npersonchoredivision}. We know how to compute a proportional allocation ( similar to the cake cutting) and an envy-free allocation \cite{envy-freechoredivisionforanyagents} for any number of agents. Though several algorithms for cake cutting also apply to chore-cutting, the theoretical properties of the two problems differ in many cases, and much less work has been done for chore-cutting than for cake cutting \cite{AlgorithmicSolutionsforEnvy-FreeCakeCutting,AlgorithmsforCdivisionofchores,complexitychoredivisio,DividingConnectedChoresfairly,ExactProceduresforenvyfreechoredivision,Chaudhury2020DividingBI}.\newline
\\
However, a fundamental issue when deploying a certain cake cutting (or, chore-cutting) algorithm  is that agents are self-interested and may manipulate and misreport their valuations  to the algorithm to get better allocations. This motivates the study of the cake-cutting problem ( or,chore-cutting) from a game-theoretic aspect, in particular 
a mechanism design aspect. Is there a \textit{truthful} and fair cake cutting mechanism such that truth-telling is each agent's dominant strategy? This question was first proposed by Chen, Lai, Parkes and Procaccia \cite{TruthJusticecakecutting}.\newline
Now one question arises: how can we represent a value density function succinctly? There are two different approaches that have been considered in the literature.
The first, in fact the most ubiquitously used model in the literature, is the Robertson-Webb model where the mechanism proceeds through a sequence of queries on the valuation functions of the agents that are of the following two types:
\begin{itemize}
    \item $\textbf{Eval}_i(x,y)$: ask agent $i$ his value on the interval $[x,y]$.
    
    \item $\textbf{Cut}_i(x,p)$: ask agent $i$ for a point $y$ where $[x,y]$ is worth exactly $p$.
\end{itemize}
Under this model, it has been shown that no truthful envy-free mechanism exists with bounded number of queries \cite{howtocutacakebeforepartyends} and that any truthful mechanism is dictatorial \cite{dictatorship} (i.e., there is a fixed agent who gets the entire cake among two agents).\\
In the second approach, the value density function is assumed to be \textit{piecewise-constant}. Piecewise constant functions are concise and may approximate general valuation functions to any accuracy \cite{howtocutacakebeforepartyends} , and they can be succinctly encoded. The mechanism then takes the $n$ encoded value density functions as inputs and outputs an allocation. These mechanisms are called \textit{direct-revelation mechanisms}.
 \\
For direct-revelation mechanisms, Chen et al. \cite{TruthJusticecakecutting} first give a deterministic truthful Pareto optimal envy-free cake cutting mechanism for any number of agents with piecewise uniform valuation functions. Chen et al.'s result show that fairness and truthfulness are compatible in the allocation of heterogeneous resources (cake). Chen et al. \cite{TruthJusticecakecutting} then proposed the following natural open problem.
\newline
\\
\textit{Does there exist a truthful envy-free cake-cutting mechanism for piecewise-constant value density functions?}
\\
\newline 
Tao \cite{OnExistenceoftruthfulfaircakecuttingmechanisms} shows the negative result of the above question raised by Chen et al.\cite{TruthJusticecakecutting}. Basically, for most questions in cake cutting there exist parallel questions in chore-cutting. Now you may think the following question :
\newline
\\
\textit{Does there exist a truthful envy-free chore-cutting mechanism for piecewise-constant value density functions?}
\newline
\\
Bei et al. \cite{Truthfulfairdivisionwithoutfreedisposal} first show the existence of a truthful envy-free chore-cutting mechanism for two agents. They also show that when each agent values a single interval of the form $[0,x_i]$,  no truthful envy-free chore-cutting mechanism exists with one of the following properties:
\begin{itemize}
    \item anonymity;

    \item connected piece assumption;

    \item position-oblivious.
    
\end{itemize}
Francis \cite{Strategyproofchoredivision} show a truthful proportional Pareto optimal algorithm for chore-cutting with piecewise uniform valuation functions. 

The purpose of this paper is to demonstrate some impossibility results regarding the existence of a truthful envy-free chore-cutting mechanism under some known constraints. 
\subsection{Our Results}
    Throughout the paper, we focus on deterministic truthful envy-free mechanisms when agents have \textit{piecewise constant valuations}. Among the several valuation function classes, this one stands out as one of the most expressive functions in the study of cake cutting \cite{cakecuttingenvyandtruth,OptimalProportionalCakeCuttingwithConnectedPieces,Optimalenvyfreecakecutting,cakecuttingalgorithmsforpiecewiseconstant,TruthJusticecakecutting,Truthfulfairdivisionwithoutfreedisposal,DeterministicStrategyproof,towardsmore}. Moreover, piecewise-constant functions become more desirable because they are concise and may approximate general valuation functions to any accuracy \cite{howtocutacakebeforepartyends}.\newline
    In the first part, a family of impossibility results related chore-cutting mechanisms is shown. Firstly, we show that no deterministic truthful envy-free mechanism exists with connected piece assumption. Even if each agent has a non-zero chore value for the whole chore i.e., $f_i(x)>0$ for all $x\in [0,1]$, the same impossible result still occurs. Secondly, we prove that no deterministic truthful dictatorship mechanism exists that is envy-free i.e., no deterministic truthful envy-free chore mechanism exists where an agent would not accept any chore valuation. After removing the dictatorial constraint, the same impossibility result also applies to non-wasteful constraints.
\subsection{Related Work}
For many years, the ``cake cutting problem" has been a significant topic in the fields of fair division and social choice. Research on fair division originated in the 1940s by the Polish mathematician Hugo Dyonizy Steinhaus to focus on protocols for achieving fairness objectives in cake-cutting. \cite{Theproblemoffairdivision}. Even though the existence and computation of fair allocations have been thoroughly researched \cite{ANENVY-FREECAKEDIVISIONPROTOCOL,Howtocutacakefairly,rental,ANoteOnCakeCutting,Stromquist1980HowTC}, Chen et al's \cite{TruthJusticecakecutting} research work that we mentioned earlier was the first to consider incentive issues. Chen et al. \cite{TruthJusticecakecutting} give the first truthful envy-free Pareto optimal cake cutting mechanism that works when each agent's valuation is piecewise uniform, a special case of piecewise constant valuation. Specifically, the mechanism proposed by chen et al.\cite{TruthJusticecakecutting} for piecewise uniform value density functions is further studied by
many researchers \cite{cakecuttingalgorithmsforpiecewiseconstant,IncentiveCompatible,TruthfulCakeCuttingForExternalities,cakecuttingenvyandtruth,Truthfulfairdivisionwithoutfreedisposal,OnExistenceoftruthfulfaircakecuttingmechanisms,DeterministicStrategyproof,truthfulcakesharing,Envyfreemechanismwithminimumnumberofcut,anenvyfreeandteruthfulmechanismswithasmallnumberofcuts,TruthfulnessPF,TruthfulFairdivision}. Aziz and Ye \cite{cakecuttingalgorithmsforpiecewiseconstant}
show that no truthful cake cutting mechanism exists if it satisfies one of the following properties:
\begin{itemize}
    \item proportional and Pareto-optimal;

    \item Robust-proportional and non-wasteful.
\end{itemize}
 Bei et al. \cite{cakecuttingenvyandtruth} show that there does not exist a truthful envy-free cake cutting mechanism under one of the four constraints:
 \begin{itemize}
     \item non-wasteful;

     \item position-oblivious;

     \item connected piece assumption;

     \item agents report their value density function functions sequentially.
     
 \end{itemize}
 Bei et al. \cite{Truthfulfairdivisionwithoutfreedisposal} show that no truthful envy-free cake cutting ( or, chore cutting) mechanism with uniform density functions of the form $[0,x_i]$ exists if it satisfies one of the four constraints:
 \begin{itemize}
     \item anonymity;

     \item connected piece assumption;
    
     \item position-oblivious.     
 \end{itemize}
Tao \cite{OnExistenceoftruthfulfaircakecuttingmechanisms} shows that there does not a exist truthful proportional cake cutting mechanism, even when all of the following hold:
\begin{itemize}
    \item there are two agents;

    \item valuation density functions are piecewise constant;

    \item each agent is hungry: for $i=1,2$, $f_i(x)>0$ $\forall x \in [0,1]$; 

    \item the mechanism needs not be entire: the mechanism may throw away parts of the cake.
\end{itemize}
 Maya and Nisan \cite{IncentiveCompatible} give a characterization of truthful and Pareto optimal mechanisms for two agents. Recently, Francis \cite{Strategyproofchoredivision} 
present a truthful proportional Pareto optimal algorithm for chore-cutting with piecewise uniform valuation functions.
\section{Preliminaries}
In a chore-cutting instance, a divisible and heterogeneous bad resource (or a ``chore''), represented by the interval\footnotemark{} $[0,1]$ to be divided among a set of agents $N=\{1,2,....,n\}$ \footnotetext{We often denote the chore by an interval $[a,b]$ for convenience. It can be normalised to $[0, 1]$.} \cite{complexitychoredivisio,DividingConnectedChoresfairly,AlgorithmsforCdivisionofchores}. Each agent $i$ is endowed with an integrable, non-negative chore density function $f_i:[0,1]\rightarrow \mathbb{R}^+\cup {0}$, which captures how the agent $i$ values different parts of the chore. By $f_i(x)$ we denote the value of density function of the agent $i$ at the point $x$.   
A \textit{piece of chore} $X$ is a finite union of disjoint subintervals of $[0,1]$. Basically, the agent $i$'s chore value on the piece $X$, denoted by $v_i(X)$, is given by 
\begin{center}
    \[v_i(X) = \sum_{I\in X}\int_{I} f_i(x) \,dx \]
\end{center}
An allocation is a partition of $[0,1]$ into a set $\{A_1,A_2,...,A_n\}$ where each $A_i$ allocated to agent $i$, and $A_i\cap A_j=\emptyset$ for any $i\neq j$. The allocation is said to be entire if  $\bigcup_{i=1}^n A_i=[0,1]$. Note that in the chore-cutting agents want to minimize their value. 
Two notions of fairness have been studied in the chore-cutting literature.\newline
An allocation $(A_1,A_2,....,A_n)$ is said to be proportional if each agent receives at most his average share of the entire chore:
\begin{center}
    $\forall i : v_i(A_i)\leq \frac{1}{n} v_i([0,1])$.
\end{center}
An allocation $(A_1,A_2,....,A_n)$ is said to be envy-free if no agent can obtain a lower share by exchanging with any other agent:
\begin{center}
    $\forall i,j: v_i(A_i)\leq v_i(A_j)$.
\end{center}
An entire envy-free allocation is always proportional. In the case of two agents, an allocation is envy-free if and only if is proportional.
\newline
A direct-revelation chore-cutting mechanism $\mathcal{M}$ is a function that takes the valuation functions $(f_1,f_2,....,f_n)$ of the agents as an input and returns an allocation $(A_1,A_2,....,A_n)$ as an output. We only consider deterministic mechanisms in this paper, meaning that the allocation is completely determined by the input density functions. Moreover, we assume that the mechanism has to allocate the entire chore to the agents.\newline
We end this section by defining a number of properties that we consider in this paper.
\newline
\textbf{Definition 1.} Let $F=(f_1,f_2,...,f_n)$ be a given vector of  input chore density functions. The indicator function  $L_{F}:{\mathbb{R}^n} \rightarrow [0,1]$ is defined as
\begin{center}
$L_F(r_1,r_2,....,r_n)=|\{x|\forall $i$, f_i(x)=r_i \}|.$ \\
where $(r_1,r_2,....,r_n)\in \mathbb{R}^n$ and $|S|$ indicates the sum total of the lengths of the intervals in $S$. 

\end{center}
\textbf{Definition 2.} A direct-revelation chore-cutting mechanism $\mathcal{M}:(f_1,f_2,....,f_n)\rightarrow (A_1,A_2,....,A_n)$ is said to satisfy 
\begin{itemize}
    \item proportionality, If it consistently provides a proportional allocation ;

    \item envy-freeness, If it consistently provides an envy-free allocation ;
   
    \item truthfulness, if each agent's dominant strategy is to report his true chore density function i.e., for each agent $i$, any $(f_1,....,f_n)$ and any $f_i^\prime$,
    \begin{center}
        $v_i(\mathcal{M}_i(f_1,....,f_n))\leq v_i(\mathcal{M}_i(f_1,...,f_{i-1},f_i^\prime,f_{i+1},...,f_n))$.
        
    \end{center}
    Note that this means misreporting a density function never benefits the agent.
    \item connected piece assumption, if each $A_i$ is always a contiguous piece.

    \item non-wastefulness, if for each agent $i$, $f_i(x)>0$ for any $x\in A_i$ i.e., no agent can receive any portion with chore value 0.

   \item Pareto optimality, if for any allocation returned by the mechanism, there does not exist any alternate allocation that makes no agent worse off and at least one agent better off with regard to the particular valuation density functions. For example suppose 
    $(A_1^\prime,A_2^\prime,....,A_n^\prime)$ is another allocation returned by $\mathcal{M}$ rather than $(A_1,A_2,....,A_n)$ with respect to the chore density functions $f_1,f_2,....,f_n$. Then $(A_1,A_2,....,A_n)$ is Pareto optimal if for all $i$, $v_1(A_i^\prime)\geq v_(A_i)$ and there exists at least one $i$ such that $v_1(A_i^\prime)> v_(A_i)$.

    \item dictatorial, if there is a fixed agent who does not accept a share with non-zero chore valuation.
    \item position oblivious, if for any two cake cutting instances with chore density functions $F=(f_1,f_2,....,f_n)$
    and $ F^\prime=(f_1^\prime,f_2^\prime,....,f_n^\prime) $ such that $L_{F}=L_{F^\prime}$, the mechanism $\mathcal{M}$ returns the allocations $\mathcal{M}(F)=(A_1,A_2,....,A_n)$ and $\mathcal{M}(F^\prime)=(A_1^\prime,A_2^\prime,....,A_n^\prime)$ such that $v_i(A_i)=v_i(A_i^\prime)$ for every $i$.
    
    \item  anonymity, if for any vectors of chore density function $(f_1,....,f_n)$ and any permutation $\sigma$ of $(1,2,...,n)$, then $v_i(A_i)=v_i(A_{\sigma^{-1}(i)}^\prime)$ for every $i$ where $\mathcal{M}(f_1,f_2,....,f_n)=(A_1,A_2,....,A_n)$ and  
    $\mathcal{M}(f_{\sigma (1)},f_{\sigma (2)},...,f_{\sigma (n)})=(A_1^\prime,A_2^\prime,....,A_n^\prime)$    
\end{itemize}
The chore mechanism is said to be dictatorial if there is an agent which never receives a piece of chore with a value above zero, no matter what the chore density functions are.\\ 
Naturally, a mechanism is obviously position-oblivious if the chore values the agents get depend only on the lengths of the pieces that different groups of agents have chore valuations on and not on where these pieces are, and is anonymous if the value that the agents get doesn't depend on their identities.\\
Note that throughout this paper we assume that the agents have piecewise-constant valuation functions. A function $f_i$ is said to be piecewise-constant if the cake can be partitioned into a finite number of intervals such that $f_i$ is constant over each interval.
\section{Impossibility Results}
\textbf{Theorem 1.} There exists no deterministic truthful envy-free mechanism that satisfies the connected piece assumption, even for two agents.\\
\newline
\textbf{Proof.} Suppose the chore is presented by $[0,9]$. Let $\epsilon>0$ be a sufficiently small constant. Consider the chore-cutting instance $F=(f_1,f_2)$ where 
\begin{equation*}
    f_1(x)=
    \begin{cases}
    1 & \text{if } x \in [3,4+\epsilon] \cup [5,6]\\
    0 & \text{otherwise}
    \end{cases}
\end{equation*}
\begin{equation*}
    \text{and}
\end{equation*}
\begin{equation*}
     f_2(x)=
    \begin{cases}
     1 & \text{if } x\in [1,2]\cup [7- \epsilon,8]  \\
     0 & \text{otherwise}
    \end{cases}.
\end{equation*}
\newline
Under the connected piece constraint, it is easy to see that both of them will get a value at least 1 and agent $2$ will get a connected piece from the left side of the chore.
\newline
Due to envy-freeness and connected piece condition, we must get
$v_2(\mathcal{M}_2(F))=1$ and $V_1(\mathcal{M}_1(F))\leq 1+\frac{\epsilon}{2}$ which implies that $[0,4-\frac{\epsilon}{4}]\subseteq \mathcal{M}_2(F)$ and $\mathcal{M}_1(F)\subseteq [4-\frac{\epsilon}{2},9]$.
Considering the scenario where agent $1$ misreports his density function to be
\begin{equation*}
    f_1^\prime(x)=
    \begin{cases}
        1 & \text{if } x \in [3,4+\epsilon]\cup [5,6] \\
        2 & \text{if } x \in [8,9] \\
        0 & \text{otherwise}
    \end{cases}
\end{equation*}
\newline
Due to envy-freeness and connected piece condition, it is also clear that agent $2$ must receive a contiguous piece from the left side of the chore $[0,9]$ and agent $1$ will receive a valuation at most $2+\frac{\epsilon}{2}$ which implies that
$\mathcal{M}_1(f_1^\prime,f_2)\subseteq [6-\frac{\epsilon}{2},9]$. Since $v_1([6-\frac{\epsilon}{2},9])=\frac{\epsilon}{2}$. This contradicts that $\mathcal{M}$ is truthful.
\\
\newline
\textbf{Theorem 2.} If every agent is required to receive a connected piece, no deterministic truthful envy-free mechanism exists, even for two hungry agents.
\\
\newline
\textit{Proof.} We will prove this theorem by contradiction. Suppose such mechanism $\mathcal{M}$ exists for two agents. We construct three chore-cutting instances and analyze the outputs of $\mathcal{M}$ on these instances. For the first two instances we show that the outputs are unique. Based on the first two instances, we show that the allocation of the last chore-cutting instance output by $\mathcal{M}$ will violate the contiguity constraint.\\
\newline
\textbf{Instance 1.} $F^{(1)}=(f_1^{(1)},f_2^{(1)})$, where $f_1^{(1)}(x)=1$ and $f_2^{(1)}(x)=1$ for $x\in [0,1]$.
\newline
Due to envy-freeness, we must have $|\mathcal{M}_1(F^{(1)})|=|\mathcal{M}_2(F^{(1)})|=\frac{1}{2}$. We will denote the allocation of $\mathcal{M}(F^{1})$ by $(A_1,A_2)$. We will use $A_1$ and $A_2$ in the the definitions of the remaining two instances.\
\newline
\textbf{Definition.} $A_1=\mathcal{M}_1(F^{(1)})$ and $A_2=\mathcal{M}_2(F^{(1)})$.
\\
\newline
Without loss of generality, we assume that $A_1=[0,.5]$ and $A_2=(.5,1]$. Since $A_1$ and $A_2$ are contiguous piece.\newline
Let $\epsilon$ be a sufficiently small positive real number.\\
\newline
\textbf{Instance 2.} $F^{(2)}=(f_1^{(2)},f_2^{(2)})$, where $f_1^{(2)}(x)=1$  for $x\in [0,1]$ and 
\begin{equation*}
    f_2^{(2)}(x)=
    \begin{cases}
        1 & \text{if } x\in A_1\\
        \epsilon & \text{if } x \in A_2
    \end{cases}.
\end{equation*}
\newline
In the following proposition we will see that the only possible output by $\mathcal{M}$ is $(A_1,A_2)$.\\
\newline
\textbf{Proposition.1.} $\mathcal{M}(F^{(2)})=(A_1,A_2)$.
\\
\newline
\textit{Proof.} Firstly, we must have $|\mathcal{M}_2(F^{(2)})|\geq \frac{1}{2}$. Otherwise, agent $1$ will receive a piece of chore of length strictly greater than $\frac{1}{2}$. Since agent $1$'s valuation function is uniform on [0,1], $\mathcal{M}$ is not envy-free.\newline
Secondly, we must have $ \mathcal{M}_2(F^{(2)}) \subseteq A_2$. Suppose that $A_2$ does not contain all of  $\mathcal{M}_2(F^{(2)})$, i.e., $|A_2\cap \mathcal{M}_2(F^{(2)})|<\frac{1}{2}$. Given that $|\mathcal{M}_2(F^{(2)})|\geq \frac{1}{2}$. 
\begin{center}
  $v_2(\mathcal{M}_2(F^{(2)}))$ 
  $=v_2(A_1 \cap \mathcal{M}_2(F^{(2)}))+v_2(A_2\cap \mathcal{M}_2(F^{(2)})) $ \\
  $=1.|A_1\cap \mathcal{M}_2(F^{(2)})|+\epsilon.|A_2 \cap \mathcal{M}_2(F^{(2)})| $ \\
  $\geq (\frac{1}{2}-|A_2\cap \mathcal{M}_2(F^{(2)})|) + \epsilon.|A_2 \cap \mathcal{M}_2(F^{(2)})|$  \\
  $=\frac{1}{2} -(1-\epsilon)|A_2 \cap \mathcal{M}_2(F_{(2)})| $\\
$> \frac{1}{2}-(1-\epsilon).\frac{1}{2}>\frac{\epsilon}{2}$.  
\end{center}
On the other hand, if agent 2 misreports his value density function to $f_2^{(1)}$ (instead of his true value density function $f_2^{(2)}$), the mechanism receives input $(f_1^{(2)},f_2^{(1)})$, which becomes instance 1 since $f_1^{(1)}=f_1^{(2)}$. In this case the allocation output is $(A_1,A_2)$, and agent 2's total value, in terms of his/her true valuation $f_2^{(2)}$, is $\frac{\epsilon}{2}$. Therefore, agent 2 can receive less value by misreporting his value density function, and $\mathcal{M}$ cannot be truthful.\newline
Putting together, we have $ \mathcal{M}_2(F^{(2)})\subseteq A_2$ and $|\mathcal{M}_2(F^{(2)})|\geq \frac{\epsilon}{2}$, which implies $M_2(F^{(2)})=A_2$. So,we must have $M_1(F^{(2)})=A_1$\\
\newline
\textbf{Instance 3.} $F^{(3)}=(f_1^{(3)},f_2^{(3)})$, where 
\begin{equation*}
    f_1^{(3)}(x)=
    \begin{cases}
        1 & \text{if } x\in A_1\\
        0.5 & \text{if } x \in A_2
    \end{cases}
\end{equation*}
\begin{equation*}
    \text{and}
\end{equation*}
\begin{equation*}
    f_2^{(3)}(x)=
    \begin{cases}
        1 & \text{if } x\in A_1\\
        \epsilon & \text{if } x \in A_2
    \end{cases}.
\end{equation*}
\\
\newline 
Now we will show in the following proposition that each agent gets exactly half from both of $A_1$ and $A_2$. Which contradicts that $\mathcal{M}$ allocates connected piece for each agents.\\
\newline
\textbf{Proposition.2.} $|\mathcal{M}_1(F^{(3)})\cap A_1|=|\mathcal{M}_1(F^{(3)})\cap A_2|=|\mathcal{M}_2(F^{(3)})\cap A_1|=|\mathcal{M}_2(F^{(3)})\cap A_2|=\frac{1}{4}$.\\
\newline
We provide a brief intuition behind the proof first. Firstly, agent $1$ cannot receive a subset of length less than $\frac{1}{2}$. Otherwise, in Instance $2$, agent $1$ will misreport his value density function from $f_1^{(2)}$ to $f_1^{(3)}$, which is more beneficial to agent $1$ ( as $f_1^{(2)}$ is uniform and agent $1$) receives a lower length by misreporting).\newline
Secondly, agent $1$ cannot receive larger than half of $A_1$. If agent $1$ receives larger than $x$, agent $1$ needs to receive less than half of $A_2$ by a length of at least $2x$ to guarantee envy-freeness. This will make the total length received by agent $1$ less than $\frac{1}{2}$.
\newline
Thirdly, agent $1$ cannot receive less than half of $A_1$. If agent receives less than half of $A_1$, agent $2$, having significantly less value on $A_2$ and high value on $A_1$. Since agent $2$ will get a piece from $A_1$ of length larger than half. This will destroy the envy-freeness of agent $1$, for that agent $2$ has take too much.\\ 
\newline
Finally, having shown that agent $1$ must receive exactly half of $A_1$, the envy-freeness of agent $1$ and proven fact that agent $1$'s received total length is at least 0.5 imply that agent $1$ has to receive exactly half of $A_2$.\\
\newline
\textit{Proof of Proposition.} First we must have $|\mathcal{M}_1(F^{(3)})|\geq \frac{1}{2}$. Suppose this is not the case: $|\mathcal{M}_1(F^{(3)})|<\frac{1}{3}$. We show that $\mathcal{M}$ cannot be truthful. Consider Instance $2$ where agent $1$'s density function is uniform. In Instance $2$, if agent $1$ misreports his/her density function to $f_1^{(3)}$, the mechanism $\mathcal{M}$ will see an input that is exactly the same as $F^{(3)}$ (notice $f_2^{(2)}=f_2^{(3)}$), and agent $1$ will receive a subset with length strictly less than $\frac{1}{2}$. However we have seen that in Proposition $1$ that agent $1$ will receive a subset with length exactly $\frac{1}{2}$ if (s)he reports truthfully. Since agent $1$'s true valuation is uniform, agent $1$ will benefit from this misreporting.
\newline
Let $|\mathcal{M}_1(F^{(3)})\cap A_1|=\frac{1}{4}+x$ where $x\in [-\frac{1}{4},\frac{1}{4}]$. Our aim is to prove that $x=0$. Agent $1$'s total utility in [0,1] is $\int_{0}^{1} f_1^{(3)}(x) dx=\frac{3}{4}$. To guarantee envy-freeness, we must have 
\begin{center}
    $v_1(\mathcal{M}_1(F^{(3)}))$=$v_1(\mathcal{M}_1(F^{(3)})\cap A_1)+v_1(\mathcal{M}_1(F^{(3)})\cap A_2)$\\
    = $1.|\mathcal{M}_1(F^{(3)})\cap A_1|+\frac{1}{2}|\mathcal{M}_1(F^{(3)})\cap A_2|$\\
    =$1.(\frac{1}{4}+x)+\frac{1}{2}.|\mathcal{M}_1(F^{(3)})\cap A_2|\leq \frac{3}{8}$      .....(1)
   
\end{center}
From (1) we get $|\mathcal{M}_1(F^{(3)})\cap A_2|\leq \frac{1}{4}-2x$.
The total length agent $1$ receives is then $|\mathcal{M}_1(F^{(3)})|\leq \frac{1}{2}-x$. Since we have seen $|\mathcal{M}_1(F^{(3)})|\geq \frac{1}{2}$ at the beginning, we have $x\leq 0$.\newline
On the other hand, since $|\mathcal{M}_1(F^{(3)})\cap A_1|= \frac{1}{4}+x$, we have $|\mathcal{M}_2(F^{(3)})\cap A_1|\geq \frac{1}{4}-x$.
Since $v_2([0,1])=\frac{\epsilon}{2}+\frac{1}{2}$ and $v_2(\mathcal{M}_2(F^{(3)})\cap A_1)=1.|\mathcal{M}_2(F^{(3)})\cap A_1|\geq \frac{1}{4}-x$, to guarantee envy-freeness for agent $2$, we must have $v_2(\mathcal{M}_2(F^{(3)})\cap A_2)\leq \frac{\epsilon}{4}+x$. Therefore $|\mathcal{M}_2(F^{(3)})\cap A_2|\leq \frac{1}{4}+\frac{x}{\epsilon}$ which implies $|\mathcal{M}_1(F^{(3)})\cap A_2|\geq \frac{1}{4}-\frac{x}{\epsilon}$.  Putting this into (1), we get
\begin{center}
   $1.(\frac{1}{4}+x)+\frac{1}{2}.(\frac{1}{4}-\frac{x}{\epsilon})\leq \frac{3}{8}$ 
\end{center}
which implies $x\geq 0$ if $\epsilon$ is sufficiently small. Therefore $x=0$, we have $|\mathcal{M}_1(F^{(3)})\cap A_1|=\frac{1}{4}$. Since agent $1$ receives exactly length $\frac{1}{4}$ on $A_1$. Due to envy-freeness, agent $1$ can get at most $\frac{1}{4}$ on $A_2$  and $|\mathcal{M}_1(F^{(3)})|\geq\frac{1}{2}$. Therefore $|\mathcal{M}_1(F^{(3)})\cap A_2|=\frac{1}{4}$.\\
\newline
\textbf{Theorem 3.} There does not exist deterministic non-wasteful truthful envy-free mechanism even with two players.\\
\newline
\textit{Proof.} Assume by contradiction that such mechanism $\mathcal{M}$ exists. Consider the chore-cutting instance with two agents whose density functions are $f_1(x)=1$ and $f_2(x)=1$ on the entire chore. The allocation $A=(A_1,A_2)$ given by $\mathcal{M}$
must satisfy $|A_1|=|A_2|=\frac{1}{2}$. Now consider another chore-cutting instance with two agents whose value density functions are
$h_2(x)=1$ and
\begin{equation*}
    h_1(x)=
    \begin{cases}
        0 & \text{if } x\in A_1\\
        1 & \text{if } x\in A_2
    \end{cases}
\end{equation*}
\newline
Due to truthfulness, agent $1$ will get the piece of chore $A_1$. Otherwise he can misreport his value density function $h_1^\prime= f_1$. Since the mechanism $\mathcal{M}$ is non-wasteful. So the agent $2$ will get the entire piece of chore $A_2$. 
Due to the envy-freeness, agent $1$ will get at most $\frac{1}{2}$ of $A_1$. This contradicts the envy-freeness of agent $2$. So such mechanism $\mathcal{M}$ does not exist.
\\
\newline
\textbf{Theorem 4.} There does not exist a deterministic truthful envy-free mechanism that is a dictatorship, even with two agents.\\
\newline
\textit{Proof.} Suppose otherwise that there is a dictatorship truthful envy-free mechanism. Consider the chore-cutting instance with two agents whose density functions are $f_1(x)=1$ and $f_2(x)=1$
on the entire chore. The allocation $A=(A_1,A_2)$ given by $\mathcal{M}$ must satisfy $|A_1|=|A_2|=\frac{1}{2}$. Consider another chore-cutting instance with two agents whose value density functions are $g_2(x)=1$ where $x\in [0,1]$ and 
\begin{equation*}
    g_1(x)=
    \begin{cases}
        0 & \text{if } x\in A_1 \\
        1 & \text{otherwise}
    \end{cases}
\end{equation*}
\newline 
For instance $(g_1,g_2)$, agent $1$ will get the $A_1$ if $\mathcal{M}$ is truthful, as otherwise she can bid $g_1^\prime(x)=1$ and get $A_1$. Moreover agent $1$ cannot get less than $A_1$ because of envy-freeness. Thus $A=(A_1,A_2)$ is the only possible allocation generated by $\mathcal{M}$  for the instance $(g_1,g_2)$.
\newline
However, by taking advantage of dictatorship condition, agent $2$ can misreport his density function and get a better allocation than $A_2$. For example agent $2$ can bid the following function $g_2^\prime$:
\begin{equation*}
 g_2^\prime =
 \begin{cases}
     \frac{1}{2} & \text{if } x\in A_1\\
     1 & \text{if } x\in A_2
 \end{cases}
\end{equation*}
\newline
Due to dictatorship, agent $1$ only gets the share from $A_1$. Without loss of generality, agent $1$ gets all of $A_1$. Due to envy-freeness, agent $2$ gets at most $\frac{3}{4}$ of $A_2$. Thus, agent $2$ will receive a strictly smaller value from manipulation , which implies that $\mathcal{M}$ cannot be truthful. 
\section{Acknowledgements}
The author would like to thank Bodhayan Roy for his suggestions which have improved the presentation of this paper significantly.

\bibliographystyle{unsrtnat}
\bibliography{references}  

\end{document}